\documentstyle[12pt]{article}
\newcommand{\half}{\textstyle{\frac{1}{2}}}

\newcommand{\beq}{\begin{equation}}
\newcommand{\beqar}{\begin{eqnarray}}
\newcommand{\eeq}[1]{\label{#1} \end{equation}}
\newcommand{\eeqar}[1]{\label{#1} \end{eqnarray}}
\newcommand{\yourabstract}[1]{
\mbox{}\\
\mbox{}\\
{\bf\noindent Abstract}\\
\begin{center}
\mbox{}\parbox[t]{5.in}{#1}
\end{center} }
\def\prd#1{{Phys. Rev. }{D#1} }

\def\pl#1{{Phys. Lett. }{#1} }

\def\rmp#1{{Rev. Mod. Phys. }{#1} }

\begin{document}
\begin{titlepage}
\begin{flushright}
CERN-TH.7431/94 \\
hep-ph/9409248 \\
September 1994
\end{flushright}
\vskip 0.8cm
\begin{center}
{\bf \large ABOUT ENTROPY AND THERMALIZATION}
\vskip 0.2cm
{\bf \large $-$ A MINIWORKSHOP PERSPECTIVE}
           \footnote{
           Work supported in part by the Heisenberg
           Programme (Deutsche Forschungsgemeinschaft).
           }
\vskip 0.6cm
{\bf Hans-Thomas Elze}$\;^a$ {\bf and Peter A. Carruthers}$\;^b$
\footnote{E-mail addresses: ELZE@CERNVM.CERN.CH,
                            CARRUTHERS@CCIT.ARIZONA.EDU}
\vskip 0.4cm
$\;^a$Theory Division, CERN, CH-1211 Geneva 23, Switzerland
\vskip 0.2cm
$\;^b$Department of Physics, Univ. of Arizona, Tucson, USA
\end{center}
\yourabstract{We present a summary and perspective view of the
Miniworkshop on ``{\it Entropy and Thermalization}''
in strong interactions (convener J. Rafelski), which was part of the
NATO Advanced Research Workshop on ``{\it Hot Hadronic Matter}''
that took place in Divonne, 27 June to 1 July 1994.}
\vskip 0.4cm
\begin{center}
{\it To appear in the Proceedings (Plenum Press).}
\end{center}
\vskip 0.3cm
\begin{flushleft}
CERN-TH.7431/94 \\
September 1994
\end{flushleft}
\end{titlepage}

${}$

\vspace{1.0cm}
\noindent
{\large {\bf ABOUT ENTROPY AND THERMALIZATION}}
\vspace{0.1cm}

\noindent
{\large {\bf $-$ A MINIWORKSHOP PERSPECTIVE}}
\vspace{1.0cm}

\hspace{2.0cm}{\bf Hans-Thomas Elze$\;^1$ and Peter A. Carruthers$\;^2$}
\vspace{0.5cm}

\hspace{2.0cm}$\;^1$Theory Division, CERN

\hspace{2.3cm}CH-1211 Geneva 23

\hspace{2.3cm}Switzerland

\hspace{2.0cm}$\;^2$Department of Physics

\hspace{2.3cm}University of Arizona

\hspace{2.3cm}Tucson, AZ 85721, USA

\vspace{1.5cm}



\noindent
{\bf INTRODUCTION}
\vspace{0.3cm}

Our intention is to summarize the main ideas brought forth in
this miniworkshop on ``{\it Entropy and Thermalization}'' in strong
interactions at high energy. In particular, some aspects and differing
views introduced during the round-table discussion, which are
not otherwise represented in these Proceedings, will be reported on here.

Anticipating our conclusions, there can be no doubt that
there exists at present a rich
diversity (if not confusion) of concepts related to the
measure of entropy to characterize high-multiplicity events in
high-energy reactions. This points to the fact that the fundamental
problem concerning the characteristic of these reactions is far
from being solved. In terms of a field theory it is still
not at all clear  how to precisely and most economically quantify their
very complex multiparticle or many degrees of freedom aspects and
relate them to experiments. Rather,
the general impression is that, at best,
one begins to see the scope of the problem and the first and still quite
conventional approaches to describe the essential
{\it disorder} of hot and dense hadronic matter, the {\it lack of
information} on high-order correlations of various kinds, and the
{\it dynamical complexity} of the underlying QCD fields. All of these
may be encoded in corresponding measures of entropy. There is a
generally shared feeling of the potential richness of collective
phenomena hiding in strong interactions at high energy, in particular
of heavy nuclei, but attempts to find an adequate formal description
to uncover them from experimental findings have mostly been modest.

In the following we try to spread the good news that our present
subject is part of one of the major scientific issues of our time,
i.e. the measure and understanding of disorder vs. order, which can be
observed to be in rapid development in several active parts of science,
and of physics
in particular. We proceed to present some still rather
divergent opinions about the essential features of entropy as well as
various first steps in the analysis of entropy production in
strong interactions.
\vspace{1.0cm}

\noindent
{\bf CAN WE MAKE SENSE OF DISORDER, LACK OF INFORMATION OR DYNAMICAL
     COMPLEXITY?} \footnote{For a related earlier discussion
     see Ref. \cite{Pete}.}
\vspace{0.3cm}

How should we assess the structures of systems that can exhibit
{\it disordered behaviour} in addition to apparently rather simple
coherent and logically
structured evolutions? By now there are many ideas about how to
approach this problem. Here we consider {\it entropy} as one useful
quantitative measure.

Sometimes apparently coherent motions such as sound waves
depend on random microscopic behaviour. In other cases a true
quantum coherence is essential. Intensity interferometry
is rooted in a field ensemble of Gaussian random variables in
the most common examples.

Curiously, physicists imagine entropy to be a gross thermodynamical
measure, determined by an integration procedure required to
convert heat transfer to a perfect differential. Chemists often
understand entropy. And computer people have a digital and
perhaps better feeling for this concept. Yet they must all be
integrated into a single framework covering such diverse aspects as
quantum limits of computation and
G\"odel's theorem
\cite{Chaitan}. The problem is to define
and find the {\it algorithm} that produces the least computational needs.
 Yet it must be capable of capturing the essential qualities of
disordered behaviour and complex structures. Surprisingly, topology
nowadays does not (yet?) play any essential role in the field
of strong interactions, which is governed
by the QCD Lagrangian methodology.

Historically, Boltzmann's genius stands out as the beacon of this
subject.
A key paradox here is related to {\it Liouville's theorem}, wherein the
many-particle entropy is conserved, contrary to what anybody
knows to be the basic issue regarding entropy.

Then,
there is {\it von Neumann's} definition of the {\it quantum entropy}
in terms of the density matrix. This is also a conserved quantity, and it
 therefore seems useless at first sight. However, the natural way out is
to consider the von Neumann entropy of intrinsically
{\it open systems} (cf. below), which generally is not a constant of
motion. This forms the basis of recent work by one of us
reported on here \cite{Thomas}.

One may also approach the problem in a
different way. This is known as {\it coarse graining},
unfortunately an all-encompassing term, one
variant known as the random-phase approximation in the initial
conditions. Another attempt to give it a precise meaning in terms
of relevant time scales in the context of particle production by
an external field is made in Ref. \cite{Jochen}.
However, there seems to be no general foundation for
such procedures, which are introduced studying individual
cases.
More surprising is that coarse graining does not distinguish the
``direction'' of time \cite{Mackey}, even though the entropy has to
increase in the process of coarse graining.
Again, this seems to be open to debate and we only want to mention
the idea that string theory may provide a natural
coarse graining by having to integrate out unobservable modes and
consequently may alter
the fundamental quantum mechanical Schr\"odinger equation in a
way that embodies an ``arrow of time'' \cite{GH}.

It is well known that
Boltzmann's H-theorem was derived from his famous kinetic
equation, based on a single-particle phase-space distribution. By
now it is clear that there are many extensions of his approach
having to do with higher-order correlations and computational
complexity among others, i.e. an infinite (quantum) hierarchy of coupled
equations and cellular automata, respectively.
Very little is known about how the latter or the
former BBGKY (and analogous Schwinger-Dyson) hierarchies can be
cast into more intuitively comprehensible
schemes. This concerns the description, for example, of
such striking phenomena as turbulence or multiparticle hadronization
processes, which are ``understood'' to some extent on the basis of
``simple'' phenomenological equations.

In particle physics we try to calculate the $S$-matrix.  From this we
calculate only probabilities of certain events. Usually the most
useful formulation is for the so-called {\it inclusive differential cross
sections}, for which selected particles in phase space are collected,
while all others are averaged over. Then, a sequence
of probabilities can be constructed and a hierarchy of
entropies follows rigorously, which are in agreement with quantum theory
despite their classical appearance.
We define, for example, the sequence of {\it inclusive probability
densities}:
\beq
\rho_1\; =\; \frac{1}{\sigma} \frac{d\sigma}{d\Gamma_1} \;\;,\;\;\;
\rho_2\; =\; \frac{1}{\sigma} \frac{d^2\sigma}{d\Gamma_1d\Gamma_2}
\;\;,\;\;\;
\rho_3\; =\; \frac{1}{\sigma} \frac{d^3\sigma}{d\Gamma_1d\Gamma_2
d\Gamma_3}
\;\;, \eeq{1}
etc., where $\Gamma_i$ denotes an appropriate phase-space variable.
Now, {\it information entropies} are generally defined by:
\beqar
S(|B) &=& - \sum_A P(A|B) \ln P(A|B) \;\;, \nonumber \\
S(|AB) &=& - \sum_C P(C|AB) \ln P(C|AB) \;\;, \nonumber \\
S(|ABC) &=& - \sum_D P(D|ABC) \ln P(D|ABC)
\;\;, \eeqar{2}
etc. Here $P(D|ABC)$ denotes the conditional probability of finding
the value $D$ of an observable
keeping values $A,B,C$ of other observables fixed, and
$\sum_D P(D|ABC)\propto\rho_3$, for example. To be truly inclusive,
each of the sets of variables $-$ $\{ A,B\}$, $\{ A,B,C\}$, etc. $-$
has to be understood to be complete; the notation is meant to
indicate especially the increasing number of exclusive variables.
These entropies are obviously closely related to experiment.
They can be reformulated in terms of {\it correlation
entropies} analogous to cumulants \cite{Pete1}, which systematically
remove irrelevant lower- order contributions.
These correlations vanish
when any variable becomes statistically independent of another.
Previous definitions of higher-order information entropy
miss this point: there, for independent distributions leading to additive
entropies, noise potentially obscures the signal of
true correlations.

Finally, {\it all probabilistic
entities can be reconstructed from the hierarchy of correlations using
generating functional techniques, and
individual events can be modelled by sampling from the
probabilities}.

Next, in order to eliminate one source of confusion, we argue that
the above introduced {\it information entropy is identical to von
Neumann's entropy if evaluated for a suitably defined open quantum
system}. To see this, we recall Eqs. (\ref{1}) defining the inclusive
densities or, rather, consider the associated probabilities
($\equiv\mbox{density}\times\mbox{flux factor}$). The same physics
of a scattering experiment, for example, can be described in a
somewhat unconventional way by calculating a partial trace (with
the exclusive variables kept fixed) of the time-evolved density matrix
of the total system and integrating over time from $-\infty$ to
$+\infty$; here the initial condition has to be specified according
to the in-state of the scattering reaction. This defines a
{\it time-independent density submatrix}, which can be diagonalized in
exclusive variables by a unitary transformation (provided these
variables correspond to quantum mechanical observables of the
system). Applying the formal results from Sect. 2 of Ref. \cite{I},
we conclude that the resulting matrix elements are indeed the
{\it probabilities} to find the corresponding values of observables
of the subsystem defined by the exclusive variables. The
``inclusive variables'' over which one averages or which are integrated
out by calculating partial traces, respectively, automatically
constitute the {\it environment} that complements the {\it open
subsystem} in the total closed system. Thus, formally, if we calculate
either information entropies according to Eqs. (\ref{2}) or the
corresponding {\it von Neumann entropies},
\beqar
S_{v.N.}(|B)&=&-\mbox{Tr}_A\;\hat{\rho}(A|B)\;\ln\hat{\rho}(A|B)
\;\;, \nonumber \\
S_{v.N.}(|AB)&=&-\mbox{Tr}_C\;\hat{\rho}(C|AB)\;\ln\hat{\rho}(C|AB)
\;\;, \eeqar{3}
etc., we obtain the same result. Here the notation parallels the one in
eq. (\ref{2}), with $\mbox{Tr}_A\hat{\rho}(A|B)\equiv\hat{\rho}(B)$
denoting a density submatrix with its elements defined on
the space associated with observable $B$, etc.
In general, the judicious choice of
the exclusive variables is dictated as much by the physical system
under consideration as the meaningful separation of subsystem and
environment, which is studied for strong interactions in Ref.
\cite{I}. Both approaches give a quantum mechanically precise
meaning to the term ``{\it coarse graining}'' by consistently
eliminating either inclusive variables or environment degrees of
freedom.

Several further points can be made. One is, What is the connection, if
any, with {\it thermodynamics}? Although not necessary, it is always an
interesting limit to consider. The merit of this limit, if justified, is
that strongly time-dependent dynamical details become irrelevant in a
stationary equilibrium state, concealing our
ignorance of the true situation. We remind the reader of the elegance
of Landau's application of relativistic fluid mechanics to
multiparticle production.

The main point is, What information is
obtained, and what does it tell us about nature?
Despite impressive advances in the precision of experimental data,
the conceptual framework for the description of
{\it multihadron production} is still deficient. Can
anything of fundamental value come out of the incredibly complicated
evolution of hadronic and nuclear collisions being analysed
with theoretical tools, which were shaped by experience with few-body
final states?

{}From the moment analysis of multiparticle correlations, we can see
interesting
and strong effects. Recent studies of Bose-Einstein correlations
suggest a new and interesting direction.
Apart from this, one can
imagine that resonance decays account for the main component of
the correlation data. This is not very fundamental.
At relativistic energies we must sift through enormous data sets in
which it is not clear that much of interest has happened.
One of the fascinating regularities is the omnipresent negative
binomial count distribution, and the associated ``linked pair''
structure of the cumulant correlations studied by one of us
(see Ref. \cite{Pete1} and references therein).

As far as entropy is concerned,
the mathematical concept related to probability theory has an intrinsic
validity not based on a particular set of variables. However,
the variables used to define the probabilities themselves
deserve more exploratory thinking in order to reveal
some essence of complex dynamical behaviour.
The subtleties of the behaviour of systems with many degrees of
freedom can defeat the methodology of $S$-matrix formulations, and
standard perturbative calculations are doomed to fail if non-linearities
are important, such as in semi-classical Yang-Mills fields
\cite{Berndt} and hadronizing QCD systems.

To conclude, we are still awaiting specific answers to the question
posed in the title of this section.
It appears challenging to study these problems
of a rather general nature and of
importance beyond the physics of strong interactions.
\vspace{1.0cm}

\noindent
{\bf ASPECTS OF ENTROPY IN STRONG INTERACTIONS}
\vspace{0.3cm}

Several members of the round table presented short contributions
highlighting their respective views on entropy and related attempts
to understand the complex irreversible behaviour in high-energy
collisions.

{\bf R. Omn\`es} asked the basic question, Given von Neumann's
definition of quantum entropy, $S=-\mbox{Tr}\hat{\rho}\ln\hat{\rho}$,
which is a constant of motion for an isolated system, what is the $S$
that increases? He argued that the answer is given by {\it decoherence
theory} in ordinary quantum mechanics and provided an outline thereof.
This work is documented in depth, for example, in the review articles
\cite{Zu0}.

The underlying reasoning, already implicitly alluded to in the above
discussion of an {\it open subsystem} and its {\it environment}, is
the following: Consider a complex dynamical system with many degrees
of freedom (e.g. $N\approx 10^{23}$), such as a piece of
solid matter. Select suitable collective observables, such as the
centre-of-mass coordinates or momenta, etc. Then, split the object
ideally into two interacting systems, $C$ and $E$, defined by the
collective (``exclusive'') and the complementing (``inclusive'')
environment degrees of freedom, respectively. The Hamiltonian
splits accordingly:
\beq
H\; =\; H_C+H_E+H_{CE}
\;\;, \eeq{4}
where the last term is responsible for energy exchange between $C$
and $E$ (``{\it dissipation}''). Starting with an initial state that is,
for example, a coherent superposition of states representing
the piece of matter located at $x_1$ and $x_2$, respectively,
and the environment in its ground state,
\beq
|\psi\rangle\; =\; a|x_1\rangle_C|0\rangle_E+b|x_2\rangle_C|0\rangle_E
\;\;, \eeq{5}
one can show that, for suitable model interactions, the environment
picks up very rapidly (due to
excessively small energy denominators) a
little excitation energy through $H_{CE}$. Most important, however,
the relevant collective subsystem density matrix obtained by tracing
over the environment degrees of freedom,
\beqar
\hat{\rho}_C(t)&\equiv &\mbox{Tr}_E\;\hat{\rho}(t)\; =\;
\mbox{Tr}_E\; |\psi (t)\rangle\langle\psi (t)| \nonumber \\
&\approx &|c_1(t)||x_1(t)\rangle\langle x_1(t)|+|c_2(t)||x_2(t)\rangle
\langle x_2(t)|
\;\;, \eeqar{6}
where $|c_1|+|c_2|=1$, becomes essentially diagonal on the same short
time scale. This is {\it environment-induced decoherence}. As far as
the collective variables are concerned, a {\it pure zero-entropy
initial state} (chosen only for simplicity) has become a {\it mixed
state with non-zero von Neumann entropy},
\beq
S_{v.N.}\; =\; -\mbox{Tr}_C\;\hat{\rho}_C\ln\hat{\rho}_C\;
=\; -(|c_1|\ln |c_1|+|c_2|\ln |c_2|)
\;\;. \eeq{7}
This effect, {\it mutatis mutandis}, is believed to be almost universal,
although explicit calculations are restricted to the class of models
that can eventually be represented by Gaussian path integrals. Some
pertinent questions are: {\it How complete} is the decoherence
effect? {\it How fast} is it? Under which conditions is the
resulting {\it entropy production irreversible}? Answers from explicit
calculations can be given, for example,
for the particular quantum mechanical systems
studied in Refs. \cite{Thomas,I}, which are constructed with an eye
on the relativistic quantum field theory extensions of decoherence
theory to be applied to strong interactions \cite{I,II}.

Note that, in the above example
of a solid piece of matter, the microscopic environment
(mostly phonons) and the macroscopic collective degrees of freedom
are known and rather clearly separated. In general, this will
not be the case. For strongly interacting hadronic systems and
high-energy collisions, in particular, the main stumbling block
preventing a deeper understanding is precisely that we know very
little about which are {\it the} relevant degrees of freedom among
an infinity of others. Single-particle
observables do not seem to provide a clue.

{\bf M. Danos} addressed the fundamental question of
{\it irreversibility in quantum physics}, which in a sense is even
prior to our discussion of entropy. He presented
a provocative point of view, which we quote directly with minor
changes \cite{Danos}:

One of the key points in dissipation in quantum physics is the
observation that time-reversal-invariant states have probability
measure zero. $-$ Generally, the physical states of a system do not
exhibit the symmetries of the Hamiltonian. This is so also for the
time-reversal symmetry. Since the Hamiltonian itself is time-reversal-
invariant, time-reversal-invariant states must exist, and indeed
they do. Only they have ``measure zero''. $-$
Rather than providing the mathematical derivation of this result
\cite{Danos1}, a more physical explanation is given here.
Take as the simplest possible
example a two-channel system. In a physical state there will be an
incoming wave in one channel, say channel 1, and outgoing waves in
both channels. The wave function (in the asymptotic region) will be
\beq
\Psi\; =\;\mbox{e}^{-ik_1x}+a\mbox{e}^{ik_1x}+b\mbox{e}^{ik_2x}
\;\;, \eeq{8}
where
$k_2^{\;2}=k_1^{\;2}-2mB$ (*),
with $B$ the inelasticity of channel 2. The time-reversed
wave function is $\Phi =\Psi^\ast$, which has amplitude- and
phase-related incoming waves in both channels, and an outgoing
wave in channel 1. To achieve that form the energy matching of
Eq. (*) must be fulfilled {\it exactly}. To actually construct
this wave would require an infinite set-up time as a consequence of
the time-energy uncertainty relation.

Hence, even though it is easy to write down an expression for the
time-reversed state of a physical state for any system, it is
principally impossible to {\it actually construct} such states. Then,
the superposition of a state and its time-reversed partner forming
a time-reversal-invariant state becomes equally impossible. Hence,
such states cannot exist in nature.

Unfortunately, we are unable to recall the spirit of the subsequent
lively discussion with the audience, which expressed
doubts about the validity of quantum mechanics, questions about the
existence of pure states and the relevance of infinite numbers of
degrees of freedom, and several others.

{\bf R. Weiner} turned the attention of those present to problems closer
related to experimental observations.
Namely, attempts to understand results of
correlation measurements and Hanbury-Brown-Twiss type interferometry
with secondary hadronic particles in terms of the space-time and
internal structure of their sources. This subject is covered in various
ways by his and others' contributions to these Proceedings, in
particular those to the Miniworkshop on ``{\it Multiparticle Dynamics}''.
We mention some interesting points concerning our present subject.

Consider a system, which is completely characterized by
its density matrix $\hat{\rho}$, in terms of {\it coherent states},
$a_k|\alpha_k\rangle =\alpha_k|\alpha_k\rangle$, where $a_k=$
annihilation operator
for mode $k$. Then, omitting the (non-trivial) sum over
modes (due to overcompleteness),
\beq
\hat{\rho}\; =\;\int d^2\alpha\; P(\alpha )|\alpha\rangle\langle\alpha |
\;\;. \eeq{10}
Quantum statistics allows the weight function to range between the
extremes of a coherent distribution and a chaotic one with $P_c(\alpha )
=\delta^2(\alpha -\alpha_0)$ and $P_{ch}(\alpha )=\pi^{-1}\bar{n}^{-1}
\exp (-|\alpha |^2/\bar{n})$, respectively. It can be shown that a
{\it chaotic distribution is necessary and sufficient to maximize the
von Neumann entropy} \cite{Weiner}.

Now, the simplest usually measured {\it Bose-Einstein correlations}
are defined by
\beq
C_2(k_1,k_2)\; =\;\frac{\rho_2(k_1,k_2)}{\rho_1(k_1)\rho_1(k_2)}\;
=\;\frac{\mbox{Tr}\;\hat{\rho}I_1I_2}{\mbox{Tr}(\hat{\rho}I_1)
\mbox{Tr}(\hat{\rho}I_2)}
\;\;, \eeq{11}
where the ``intensities'' are given by number operators,
$I_j=I(k_j)=a^+_{k_j}a_{k_j}$, and the densities follow from Eqs.
(\ref{1}) with $\Gamma =k$. Thus, in principle, Bose-Einstein
correlations measure the density matrix of the system and provide a
test for randomization. The very existence of non-trivial correlations,
i.e. $C_2(k_1,k_2)>1$, which is observed in a wide range of experiments,
 is {\it evidence for a (partial) randomization and non-vanishing
entropy}. However, unfortunately this aspect is usually taken more or
less for granted and, so far, empirical parametrizations of these
correlations are only employed to derive source geometry and lifetime
parameters.

{\bf U. Heinz}, finally, recalled the
apparently strongly disordered outgoing state
in a heavy-ion collision. Then,
Why is there a $\beta =T^{-1}$ characterizing the
exponential slope of major parts of the spectra of secondaries?
There are partial answers to this decades-old question,
but one still feels a lack of understanding: $T$ being a Lagrange
multiplier for the variational problem to {\it maximize} $S$ {\it at
constant energy}, one asks oneself, Where does the measured large value
of this entropy come from? Given that a local thermodynamic
equilibrium description of high energy density matter in a collision
can be justified, which in turn is still only partly understood
\cite{Klaus}, the entropy in terms of particle multiplicities is an
important additional piece of information to
distinguish various phenomenological equations of state
\cite{Letessier,Tounsi}.
\vspace{1.0cm}

\noindent
{\bf WHAT DO NEXT?}
\vspace{0.3cm}

Instead of providing {\it the} answer to Lenin's question
for our present subject, we discuss the contributions to
this miniworkshop, with more details to be found elsewhere in
these Proceedings.

The most recent and surprising results of the most conservative
approach to entropy in heavy-ion collisions, in the sense that it has
already a history of four decades, were described by {\bf J. Letessier}
\cite{Letessier}. It is based on the hypothesis of the formation of
a {\it fireball}, i.e. a space-time region of hot and
dense hadronic matter with approximately thermal properties in
the centre of mass of these reactions. Its total energy $E$ and baryon
number $B$ are assumed to be fixed and their internal properties
(particle content, phase-space distributions, etc.) to be determined
consistently by a {\it kinetic equilibrium temperature} $T$ and
{\it fugacities} $\lambda_i=\exp\mu_i/T$ of the constituents (as well as
additional parameters). The major advantage of this model is its
conceptual and technical {\it simplicity}, in principle, which allows
direct comparison with experimental results on multiplicities.
Here the measured final state {\it specific entropy}
$S/B$ is employed to discriminate between different equations
of state. It is argued that a {\it hot hadronic gas scenario is
unable to fit all available data from the 200 GeV A CERN experiments,
whereas incorporating a temporarily existing quark-gluon plasma
gives a satisfactory description} (``too good to be true'')
\cite{Letessier}.

The success of this approach crucially depends on the assumptions
of (local) thermal
equilibrium joined with a simple ``macroscopic'' collective
expansion of the fireball. One further important {\it consistency} check
should be to study two-particle correlations in precisely the same model,
as well as single-particle spectra \cite{Uli}.
Finally, it is stated that $\approx 70\%$ {\it of
the measured entropy must be produced already during the pre-thermal
phase}, the study of which thus becomes crucial also for the
understanding of the fireball model parameters.

The above and the following model share the uncertainty (``flexibility'')
 about details of the {\it hadronization} from the parton phase.
According to folklore, there is essentially
{\it no entropy production} during the (non-equilibrium) phase
transition. This is important for the interpretation, e.g. in a
thermal model, of
\beq
\frac{dS}{dy}\; =\; c_{qg}\frac{dN_{qg}}{dy}|_{b=0}\;\propto\;
\frac{dN_\pi}{dy}|_{b=0}
\;\;, \eeq{12}
which is a typical relation
between the entropy density calculated in a partonic
model, see below, and the observed hadron multiplicity.

A study of the early space-time evolution was presented by
{\bf K. Geiger} \cite{Klaus} employing a {\it parton cascade} approach
to simulate quark and gluon transport during hadronic or nuclear
collisions. This probabilistic scheme is based on state-of-the-art
{\it perturbative QCD} cross sections, which are employed similarly
as in simulations of Boltzmann equations including collision terms.
Partons are sampled from measured {\it structure functions} and
{\it propagated classically} in accordance with Altarelli-Parisi
type equations. A justification of this procedure for multiple
scatterings in extended dense systems, i.e. multiple inter-cascade
interactions, seems rather difficult within the parton model, since it
goes beyond proved factorization theorems.

Here the total {\it entropy} arises from three contributions,
\beq
S\; =\; S_{primary}+S_{secondary}+S_{hadronization}
\;\;. \eeq{13}
The {\it primary contribution}, which amounts to about 40\% of the
total, stems from the {\it decoherence process} that sets in once
the incoming hadronic wave functions are perturbed by initial-state
QCD interactions. This is the dynamical origin of the structure
functions, which has recently been addressed in Refs. \cite{Thomas,I,II}.
 Another way of stating this is by
recalling the definition of structure functions as being related to
{\it inclusive cross sections}; according to our discussion
following Eq. (\ref{1}), there must be an associated entropy.
At present, this contribution {\it cannot} be calculated {\it ab initio}
or in any quantitative model.

The {\it secondary contribution}, which accounts for practically
all the rest of the produced entropy, is due to the
{\it production of secondary partons} in elementary bremsstrahlung
or scattering processes. This is essentially analogous to what
happens in any kind of molecular dynamics simulation, namely a
covering of the available classical single-particle phase space via
scattering. Finally, the {\it hadronization contribution} is
arguably considered to be small and taken into account by hadronization
prescriptions based on universal parton-hadron duality and fitted,
for example, to $e^+e^-$ data.

The most interesting result in the present context is the {\it rapid
saturation of entropy production} (together with a thermalization of
parton spectra) on a time scale of 0.5 fm/c or less, with a value of
the specific entropy per particle, $S/N\approx 4$, which is more or
less the ideal parton gas value \cite{Klaus}.

The {\it Schwinger mechanism}, i.e. $e^+e^-$ creation in a
time-independent homogeneous electric field, was studied by {\bf J.
Rau} \cite{Jochen} w.r.t. entropy production and irreversibility.
The main result is that the {\it ``relevant'' entropy} defined here
in terms of the single-particle ($e^\pm$) occupation numbers,
\beq
S_{rel}(t)\; =\;\sum_{all\; modes}\left\{\half n_-\ln\half n_-
+(1-\half n_-)\ln (1-\half n_-)+[\; n_-\leftrightarrow n_+\; ]
\right\}
\;\;, \eeq{14}
tends to increase. However, there are two essential time scales for
the process: the {\it memory time}, $\tau_{mem}\approx (\hbar /m)
+(m/qE)$, and the {\it production time}, $\tau_{prod}\approx (m/qE)
\exp (\pi m^2/2\hbar qE)$. Depending on their relative size the
process is essentially Markovian and {\it irreversible} (weak fields),
leading to monotonically increasing $S_{rel}$, or else it shows
important {\it memory effects} (strong fields), leading to oscillations
of $S_{rel}$ on the scale of $\tau_{mem}\approx\hbar /m$ \cite{Jochen}.
These effects are analogous to what happens with the Boltzmann
equation depending on the relative size of the time between collisions
(``$\tau_{mem}$'') and the duration of a single collision
(``$\tau_{prod}$'').

It seems important to realize how the ``relevant'' entropy here fits
into our preceding discussion of coarse graining, inclusive variables,
and open systems with their environments. Clearly, the
entropy $S_{rel}$ determined by occupation numbers is {\it relevant}
w.r.t. experiments measuring single-particle observables. However,
it corresponds to a {\it chosen cut in the space of observables} of
the system. Thus, one deliberately discards information in an
inclusive way, e.g. about relative phases of outgoing single-particle
waves or higher-order correlations ($n$-point functions with $n>2$),
which amounts to a coarse graining and results in information entropy as
before. In distinction, the considerations in Refs. \cite{Thomas,GH,I,
Zu0} are based on the observation that in some systems or theories
(e.g. QCD) there is a {\it dynamical cut in the space of fundamental
modes} of the system, which naturally separates it into an
``observable'' subsystem and its environment. In this case, the coarse
graining is dictated by the complex system itself. Then, the {\it von
Neumann entropy} related to all information available about the
subsystem is {\it not} a constant of motion and is the relevant
entropy. An {\it additional} coarse graining, such as a restriction
to inclusive single-particle observables, may still be necessary for
practical purposes.

The contribution by {\bf H.-Th. Elze} \cite{Thomas}
provides a simple introduction to the mechanism of {\it
environment-induced quantum decoherence} (cf. the above discussion
of R. Omn\`es' presentation) with a view towards
strong interactions. In QCD a separation of
non-perturbatively interacting, almost constant, field configurations,
which can neither hadronize nor initiate hard scatterings, from the
usual high-energy or far off-shell partons seems essential to attack
the strong-coupling problem underlying entropy production in
multiparticle processes.

We mention two particular results. Employing the {\it Schmidt
decomposition} of the complex system density matrix, see Sect. 2 of
Ref. \cite{I}, one finds that the von Neumann entropy for the
subsystem always equals the one for its environment. Therefore,
one may choose to {\it eliminate
either the environment or the subsystem} degrees of freedom,
whichever is simpler. Secondly, in the example of the inverted
oscillator \cite{Thomas}, which is partially chaotic in the classical
limit, one observes an {\it exponentially growing entropy production},
which is governed essentially by the classical Lyapunov exponent.
Here, the decoherence is induced by the coupling to
the {\it vacuum fluctuations} of only one environment oscillator. Thus,
under suitable conditions an extremely simple zero-temperature
environment is sufficient to cause entropy production in the
subsystem, which might be relevant in the following.

The work on {\it chaos
and entropy production in classical Yang-Mills fields} reported by
{\bf B. M\"uller} \cite{Berndt} addresses the question of entropy
production as being connected intimately to the problem of thermalization
 in strongly interacting systems. One studies the chaotic time
evolution of classical Yang-Mills fields employing the lattice gauge
theory discretization for the Hamiltonian equations of motion. Thus,
it is shown that a random {\it ensemble of initial field configurations
self-thermalizes} rapidly, i.e. the probability distribution of the
magnetic plaquette energy evolves into an exponential Boltzmann
distribution. Furthermore, the maximal Lyapunov exponent of the
time-dependent classical system
is demonstrated to yield the {\it damping rate of coloured collective
(plasmon) excitations} $\propto g^2T$, which is calculated otherwise
by finite-temperature QCD perturbation theory ($T\gg T_c$). Its
inverse yields a {\it thermalization time}, which
rapidly decreases with increasing $T$ from $\tau_S^0\approx 0.5$ fm/c at
$T\approx 200$ MeV. Employing the complete Lyapunov spectrum,
which presumably corresponds to including other unstable collective
excitations at finite $T$, an even shorter thermalization time can
be deduced from the rate of entropy growth, $\tau_S=\bar{S}_{equil}/
\partial_t\bar{S}$. Herein, the relevant entropy is the
{\it Kolmogorov-Sinai entropy} $\bar{S}$, which arises by a coarse
graining of the classical phase space \cite{Berndt}.

Two related points seem to deserve further study in order to fully
understand these remarkable results. Firstly, where does the
ensemble of initial field configurations come from? Following Refs.
\cite{Thomas,I}, one is led to conjecture that the {\it environment
of high-energy or far off-shell partons} and integrating out these
ultraviolet degrees of freedom, respectively, result in the effectively
{\it classical} initial conditions above. Secondly, are the strongly
coupled Yang-Mills system under consideration and its evolution stable
w.r.t. the ultraviolet quantum fluctuations? Here, {\it asymptotic
freedom} may help to keep such stability, which is necessary
in order to relate this approach to actual hadronic or nuclear
collisions.

In conclusion, we hope to have raised or rephrased some
interesting questions to stimulate further research on entropy and
thermalization, particularly in strong interactions. We thank all
participants of the Miniworkshop for sending copies of
their presentations and, especially, J. Rafelski for the intellectual
and organizational support without which it would not have happened.


\begin{thebibliography}{15}
\bibitem{Pete} P. A. Carruthers, in Proc. NASI ``Hot and Dense Nuclear
               Matter'', Bodrum (Turkey),
               1993, to be published by Plenum Press.
\bibitem{Chaitan} R. Landauer, Phys. Today 45, No. 5 (1991) 23; \\
                  C. J. Chaitan,
                  ``Information, Randomness and Incompleteness'',
                  World Scientific, Singapore, 1987.
\bibitem{Thomas} H.-Th. Elze, contribution to these Proceedings.
\bibitem{Jochen} J. Rau, contribution to these Proceedings.
\bibitem{Mackey} M. C. Mackey, ``Time's Arrow: The Origins of
                 Thermodynamic Behaviour'', Springer Verlag, New York,
                 1992.
\bibitem{GH} J. Ellis, N. E. Mavromatos and D. V. Nanopoulos,
             \pl{B293} (1992) 37; preprint CERN-TH.7195/94;
             and references therein; \\
             M. Gell-Mann and J. B. Hartle,
             \prd{47} (1993) 3345.
\bibitem{Pete1} P. A. Carruthers, Int. J. Mod. Phys. A4 (1989) 5587.
\bibitem{I} H.-Th. Elze, preprint CERN-TH.7131/93 (hep-ph/9404215),
            to appear in Nucl. Phys. B.
\bibitem{Berndt} B. M\"uller, contribution to these Proceedings.
\bibitem{Zu0} W. H. Zurek, Phys. Today 44, No. 10 (1991) 36; \\
              R. Omn\`es, \rmp{64} (1992) 339.
\bibitem{II} H.-Th. Elze, preprint
            CERN-TH.7297/94
            (hep-th/9406085), submitted to Phys. Rev. Lett.
\bibitem{Danos} M. Danos, private communication.
\bibitem{Danos1} M. Danos, NIST Technical Note 1403 (1993).
\bibitem{Weiner} R. Weiner, private communication.
\bibitem{Klaus} K. Geiger, contribution to these Proceedings.
\bibitem{Letessier} J. Letessier, contribution to these Proceedings.
\bibitem{Tounsi} A. Tounsi, contribution to these Proceedings.
\bibitem{Uli} U. Heinz, contribution to these Proceedings.
\end{thebibliography}
\end{document}